\documentclass{PoS}

\newcommand{\beq}{\begin{eqnarray}}
\newcommand{\eeq}{\end{eqnarray}}
\usepackage{subfigure}

\title{Search for the IR fixed point in the Twisted Polyakov Loop scheme (II)}

\ShortTitle{Search for the IR fixed point}

\author{ \speaker{Etsuko Itou}$^a$, Tatsumi Aoyama$^b$, Masafumi Kurachi$^{b}$, C.-J. David Lin$^c$, Hideo Matsufuru$^d$, Hiroshi Ohki$^{b}$, Tetsuya Onogi$^a$, Eigo Shintani$^e$ and Takeshi~Yamazaki$^b$\\
$(a)$Department of Physics, Osaka University, Toyonaka 560-0043, Japan \\
$(b)$Kobayashi-Maskawa Institute for the Origin of Particles and the Universe, Nagoya University, Nagoya, Aichi 464-8602 Japan\\
$(c)$Institute of Physics, National Chiao-Tung University, and National Center for Theoretical Sciences, Hsinchu 300, Taiwan\\
$(d)$High Energy Accelerator Research Organization (KEK), Tsukuba 305-0801, Japan\\
$(e)$RIKEN-BNL Research Center, Brookhaven National Laboratory, Upton, NY 11973, USA \\
E-mail: \email{itou@het.phys.sci.osaka-u.ac.jp}
}

\abstract{We measure the renormalized coupling in the Twisted Polyakov loop scheme for SU(3) gauge theory coupled with $N_f=12$ fundamental fermions.
To find the infrared fixed point of this theory, we focus on the step scaling function for the renormalized coupling.
We take the continuum limit using the linear function of $(a/L)^2$ and a constant fit function.
We find that there is a sizeable systematic error due to the choice of the continuum extrapolation function in the low energy region.
We will give some directions to reduce the systematic errors.}

\FullConference{The XXVIII International Symposium on Lattice Field Theory, Lattice2010\\
		June 14-19, 2010\\
		Villasimius, Italy}

\begin{document}

\section{Introduction}
Discovering the fixed point and the critical phenomena around it are one of the intriguing subjects in quantum field theory.
The fixed point is defined by the zero point of the beta function of the theory, and there the theory exhibits scale invariance. Some of them are exactly solvable.
The values of the coupling constant at the fixed point depend on the renormalization scheme.
In general, changing the renormalization scheme corresponds to the coordinate transformation in theory space.
Thus, the renormalization group flows are changed by this transformation, but the existence of fixed point and whose critical exponents are not changed\cite{Wilson:1973jj,Sumi:2000xp}.

Recently, there have been many papers concerning the fixed point search and the study of phase structure of SU(N) gauge theory with a large number of flavors.
The perturbative $\beta$ function indicates the existence of non-trivial infrared (IR) 
fixed points for a certain range of the number of large-flavor ($N_f$) SU(N) gauge theories, which is so-called ``conformal window".
Possible appearance of these IR fixed points has stimulated phenomenological studies of topics such as
dynamical electro-weak symmetry breaking and unparticle physics.

The existence of these IR fixed points depends on the gauge group, the number of flavors, and the representation 
of fermion fields.
For SU(3) gauge theory with fermions in the fundamental representation, such a fixed point has been 
predicted in the range $8< N_f \le 16$ using 
perturbation theory \cite{Banks:1981nn}.
However, the value of the renormalized coupling may be in
the regime where perturbation theory is not applicable.
There are some analytical studies with improvements of the perturbation for the search for the lower bound of the conformal window, and the largest one give a prediction that $N_f \le 12$ is out of the window\cite{Gardi:1998rf}.

One of the nonperturbative approaches to this subject is a lattice simulation, and there are some recent studies concerning it.
First such lattice study for SU(3) gauge theory was carried out in Ref.~\cite{Damgaard:1997ut}, where the authors investigated 
the phase structure of the case of $N_f=16$.
Appelquist {\it et al.} performed lattice calculation of the running coupling constant in the 
Schr\"{o}dinger functional (SF) scheme and discovered evidence of an IR fixed point in the case of 
$N_{f} = 12$~\cite{Appelquist:2009ty}.
On the other hand, Fodor {\it et al.} does not obtain a signal of IR fixed point using the different scheme based on the Wilson loop\cite{Bilgici:2009kh, Fodor:2009rb}.
The difficulty is mainly due to scheme dependence of the running coupling constant and the presence of significant
lattice artifacts in the strong-coupling regime.
Therefore it is important to measure the running coupling in different renormalization schemes, and estimate the dicretization error carefully.

In this work, we perform lattice simulation of the running coupling constant for the fundamental representation,
$N_f = 12$, SU(3) gauge theory.
Similar to the approach of Appelquist {\it et al.}, we measure the step scaling function $\sigma (s, g^2(L))=g^2(sL)$ 
keeping the values of bare coupling constant ($\beta$) that give constant renormalzed coupling ($g^2(L)$) for each small lattice size.
We work in the Twisted Polyakov Loop (TPL) scheme which does not contain $O(a/L)$ discretization errors.
This TPL scheme was first proposed by de Divitiis {\it et al.} \cite{deDivitiis:1993hj,deDivitiis:1994yp} for SU(2) gauge theory,
and we extend the definition of the scheme to the SU(3) case\cite{Bilgici:2009nm}.

In this paper, we give a short review of TPL scheme in \S.\ref{sec:TPL}. 
Our preliminary results for $N_f=12$ SU(3) gauge theory is reported in \S.\ref{sec:Nf-12}.
In the last section, we discuss some ways of improvements to reduce the discretization error and show the future direction to search for the IR fixed point.

\section{Twisted Polyakov loop (TPL) scheme}\label{sec:TPL}
In this section, we present the definition of the Twisted Polykov Loop scheme in SU(3) gauge theory.
This is an extension of the SU(2) case as discussed in Ref.~\cite{deDivitiis:1993hj}.
To define the TPL scheme, we introduce twisted boundary condition for the link variables in $x$ and $y$ directions on the lattice:
\beq
U_{\mu}(x+\hat{\nu}L/a)=\Omega_{\nu} U_{\mu}(x) \Omega^{\dag}_{\nu}. \hspace{0.5cm}  (\nu=x,y) \label{twisted-bc-gauge}
\eeq
Here, $\Omega_{\nu}$ are the twist matrices.% which have the following properties:
%\beq
%\Omega_{x}\Omega_{y}=e^{i2\pi/3}\Omega_{y}\Omega_{x},\quad
%\Omega_{\mu} \Omega_{\mu}^{\dag}=1, \quad
%(\Omega_{\mu})^3=1, \quad
%\mbox{Tr}[\Omega_{\mu}]=0.
%\eeq
The gauge transformation for link variables $U_\mu (r) \rightarrow \Lambda (r) U_\mu (r) \Lambda^\dag (r+\hat{\mu})$ and eq.(\ref{twisted-bc-gauge}) imply 
the gauge transformation at boundary with a twisted gauge matrix:
\beq
\Lambda (r+ \hat{\nu}L/a)=\Omega_\nu \Lambda(r) \Omega_\nu^\dag.
\eeq
Because of this twisted boundary condition, the definition of Polyakov loops in the twisted directions are modified,
\beq
P_{x}(y,z,t) ={\mbox{Tr}} \left( [ \prod_{j} U_{x}(x=j,y,z,t)] \Omega_{x} e^{i2\pi y/3L} \right),
\eeq
in order to satisfy gauge invariance and translation invariance, and similary for $y$-direction.

The renormalized coupling in TPL scheme is defined by taking 
the ratio of Polykov loop correlators in the twisted ($P_x$) and the untwisted ($P_z$) directions:
\beq
g^2_{TP}=\frac{1}{k} \frac{\langle \sum_{y,z} P_{x} (y,z,L/2a) P_{x} (0,0,0)^{\dag} \rangle}{ \langle \sum_{x,y} P_{z} (x,y,L/2a) P_{z} (0,0,0)^{\dag} \rangle }.\label{TPL-def}
\eeq 
At tree level, this ratio of Polyakov loops is proportional to the bare coupling.  The
proportionality factor $k$ is obtained by analytically calculating the one-gluon-exchange diagram.
To perform this analytic calculation, we choose the explicit form of the twist matrices~\cite{Trottier:2001vj},
%\beq
%\Omega_x=\left( 
%\begin{array}{ccc}
%0 & 1 & 0\\
%0 & 0 & 1\\
%1 & 0 & 0
%\end{array}
%\right),
%\Omega_y=\left( 
%\begin{array}{ccc}
%e^{-i2\pi /3} & 0 & 0\\
%0 & e^{i2\pi /3} & 0\\
%0 & 0 & 1
%\end{array}
%\right).
%\eeq 
and in the case of SU(3) gauge group we found $k\sim 0.03184$~\cite{Bilgici:2009nm}.

%The naive twisted boundary condition for lattice fermions can be written by
%\beq
%\psi (x+\hat{\nu}L/a)=\Omega_{\nu} \psi(x).
%\eeq
%However, this results in an inconsistency when changing the order of translations, namely,
%\beq
%\psi (x+\hat{\nu}L/a+\hat{\rho}L/a)&=&\Omega_{\rho} \Omega_{\nu} \psi(x), \nonumber\\
%&\ne &\Omega_{\nu} \Omega_{\rho} \psi(x).
%\eeq  
To introduce the fermions which satisfy both the twisted boundary condition and translation invariance on the lattice, we have to introduce additional ``smell" degree of freedom and indentify the fermion field as a $N_c \times N_s$ matrix ($\psi^a_\alpha$(x)), where $N_c$ and $N_s$ are the numbers of color and smell degrees of freedom respectively~\cite{Parisi:1984cy}.
Then we impose the twisted boundary condition for fermion fields to be 
\beq
\psi^a_{\alpha} (x+\hat{\nu}L/a)= e^{i \pi/3} \Omega_{\nu}^{ab} \psi^{b}_{\beta} (\Omega_{\nu})^\dag_{\beta \alpha}
\eeq
for $\nu=x,y$ directions.
Here, the smell index can be considered as a ``flavor'' index, then the number of flavors should be a multiple of $N_s(=N_c=3)$.
We use staggered fermion in our simulation. This contains four tastes for each flavour.
This enables us to perform simulations with a multiple of $12$-flovors in this SU(3) gauge theory with twisted boundary condition.

At the end of this section, we would like to remark on the center symmetry of this theory.
The generator of the symmetry is given by $z=exp(2 \pi i k/3)$, $k=0,1,2$ for $SU(3)$ gauge theory.
Although the Wilson gauge action is invariant under the following transformation for the link variable for each direction,
\beq
U_\mu(t,\vec{x}) \rightarrow z U_\mu(t,\vec{x}),
\eeq
the fermion action is not invariant.
Therefore the vacuum expectation values of operator at $z=1$ and at $z=exp (\pm 2\pi i/3 )$ are different.
In the simulation, we generated the gauge configurations at nontrivial vacua where the vacuum expectation value of Polyakov loop correlator in untwisted direction has a nontrivial phase.
We also investigated the one-loop effective potential for each vacuum analytically, and found the nontrivial one is the true vacuum\cite{SU3-final-paper}.

\section{Simulation detail}\label{sec:Nf-12}
\subsection{Simulation set up}
The gauge configurations are generated by the Hybrid Monte Carlo algorithm, and we use the Wilson gauge and the staggered fermion action.
To reduce large statistical fluctuation of the TPL coupling, as reported
in Ref.~\cite{Divitiis:1995}, 
we measure Polyakov loops at every Langivin step and perform 
a jackknife analysis with large bin size, typically of $O(10^3)$.
This enables us to evaluate the statistical error correctly.
The simulations are carried out with several lattice sizes ($L/a=4,6,8,10,12,16$)
at more than twenty $\beta$ values in the range $4.5 \leq \beta \leq 20$.
We generate 50,000--100,000 configurations for each $(\beta, L/a)$ combination.

\begin{figure}[htbp]
% \begin{minipage}{0.5\hsize}
  \begin{center}
\rotatebox{0}{
   \includegraphics[width=6cm,clip]{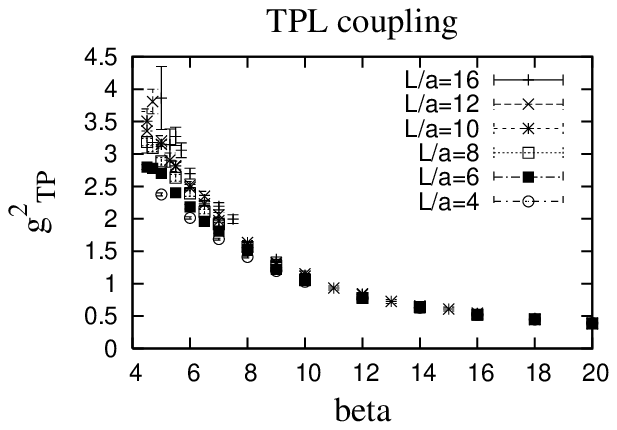}
\quad \quad
  \includegraphics[width=5.3cm,clip]{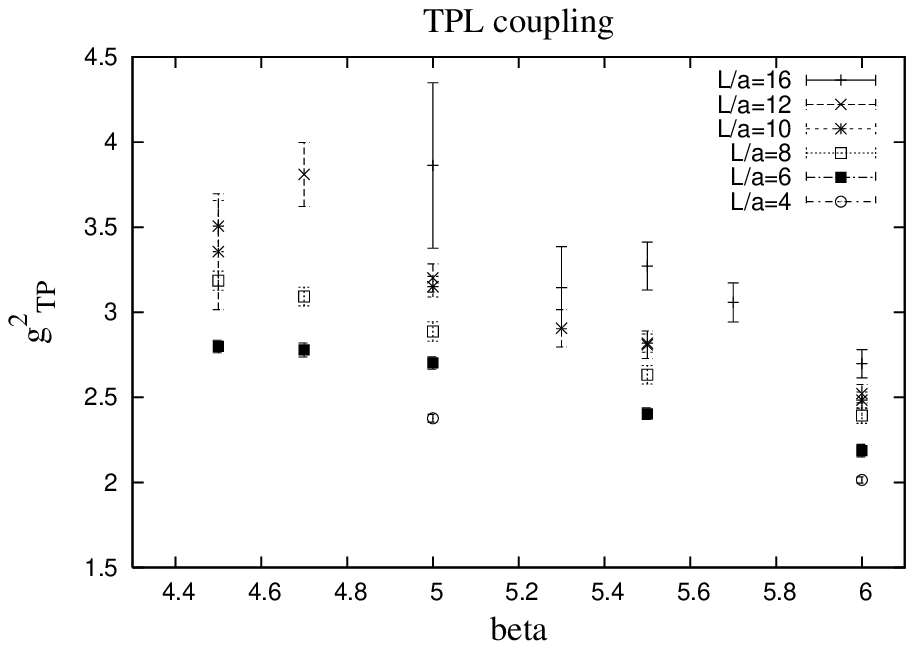}
}
  \end{center}

  \caption{TPL renormalized coupling in the each $\beta$ and $L/a$ in $N_f=12$. The right figure shows the detail behavior in low $\beta$ region.} 
  \label{fig:global-fit}

\end{figure}

Figure~\ref{fig:global-fit} shows the $\beta$ dependence of the 
renormalized coupling in TPL scheme at various lattice sizes.
The results can be fitted at each fixed lattice size 
to the interpolating function
\begin{equation}
g^2_{\mathrm{TP}}(\beta) = \frac{C_1}{\beta}+\frac{C_2}{\beta^2}+\frac{C_3}{\beta^3}+\frac{C_4}{\beta^4},
\end{equation}
where $C_i$ are the fit parameters.

The right figure in Fig.~\ref{fig:global-fit} shows the detailed behavor of low-beta region.
We find that $g_{TP}^2$ increases monotonously with the increase of the lattice sizes within the whole range of $\beta$ examined.
This holds even at $\beta=4.5$
This is in the contrast to the case of Schor\"{o}dinger functional scheme where the renormalized coupling exhibits some ``crossing" behavior~\cite{Appelquist:2009ty}.

\subsection{The continuum extrapolation and the scaling function}
We investigate the growth rate of TPL coupling in the continuum limit within the step scaling method.
The procedure of the step scaling is first to find a set of bare coupling constant ($\beta$) for each small lattice size, which gives an input value of renormalized coupling ($g^2(\beta,a/L)=u$).
Next, we measure the step scaling function $\Sigma (u,s,a/sL)=g^2(\beta,sL/a)|_{g^2(\beta,a/L)=u}$.
Finally, we take the continuum limit and obtain the step scaling function $\sigma (s,u)$ in the continuum.
\beq
\sigma (s,u)= \lim_{a \rightarrow 0} \Sigma(u, s, a/sL)|_{g^2(L)=u}.
\eeq
In our study, we use $L/a=4,5,6,8$ as a small lattice size of the step scaling, and estimate the coupling constant for $L/a=5$ from interpolations at the fixed $\beta$ using the interpolation fit results of the lattice sizes $L/a=4,6,8$.
The step scaling parameter is $s=2$, thus a large lattice size of the step scaling is $L/a=8,10,12,16$.
From now, we denote the step scaling function $\sigma(s=2, u)\equiv \sigma(u)$.

The total error of the step scaling function can be estimated by the sum of the statistical error from each data and the systematic errors.
The systematic errors have two origins.
One of them is included in each data value $\Sigma(u,s,a/sL)$, in which the value of $\beta$ that is tuned to giving an input $u$ have the error.
The other one comes from the continuum extrapolation.
If we measure the running behavior of coupling constant, the systematic error which comes from $\beta$-tuning is accumulated.
However, we can carry out each step-scaling procedure independently for a given $u$. 
We focus on the growth rate of the renormalized coupling for several values of $u$.
And, if the running coupling constant reaches the fixed point, the growth rate $\sigma(u)/u$ should be $1$.

Now, we have to consider the statistical error and the systematic error which come from the continuum extrapolation.
To estimate the systematic error, we take the continuum limit using both a constant fit and a linear function in $(a/L)^2$.
\begin{figure}[h]
\begin{center}
\rotatebox{0}{
   \includegraphics[width=6cm,clip]{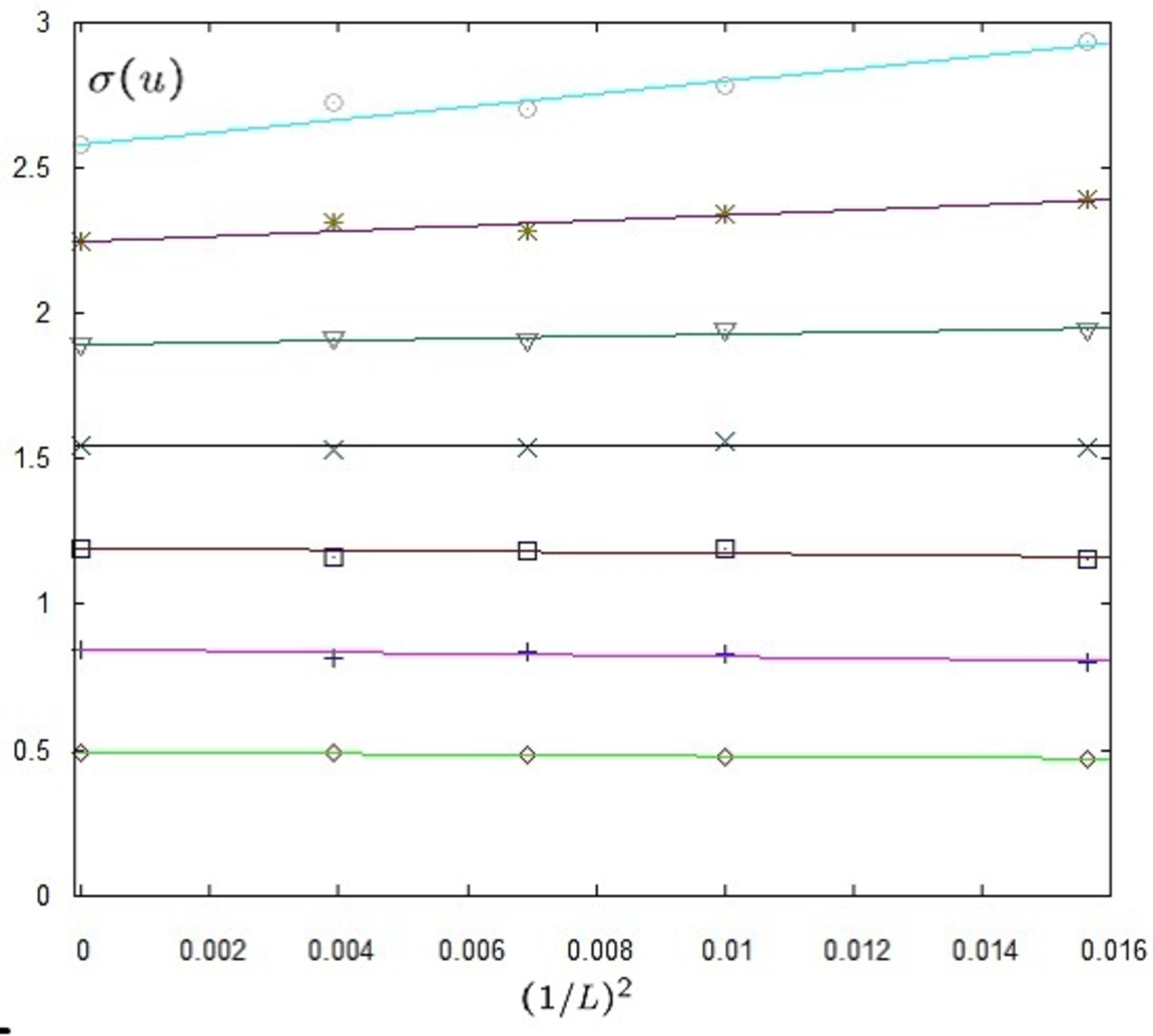}
\quad \quad
  \includegraphics[width=5cm,clip]{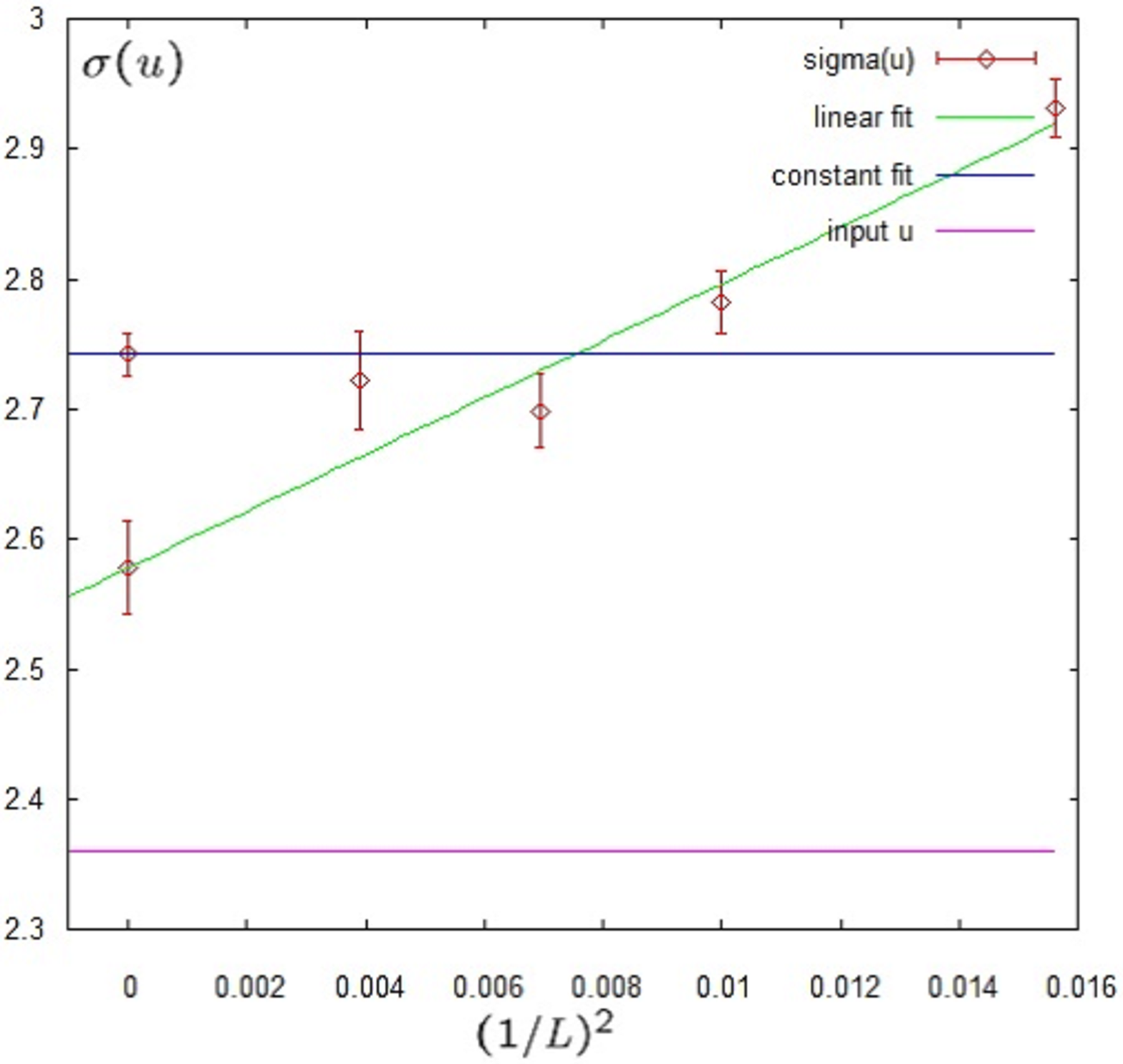}
}
\end{center}
\caption{The continuum limit of $g^2_{TP}$ with $s=2$. In the left figure, each line denotes a linear fit function of $(a/L)^2$. The statistical error bars are of the same size of the symbols (less than $3\%$). The each input value of TPL coupling ($u=g^2_{TP}(L/a)$) is $u=2.36,2.04,1.73,1.41,1.10,0.79,0.47$ from top to bottom. The right figure shows the detailed behavior of the lowest energy scale in the left figure.
The pink line denotes the value of the input renormalized coupling $u=2.36$. 
The green and blue lines denote the linear extrapolation for four points ($L/a=8,10,12,16$) and the constant extrapolation for three points ($L/a=10,12,16$) respectively.}
\label{fig:cont-lim}
\end{figure}
In the case of quenched QCD, we found that the coupling constant of the TPL scheme exhibits scaling behavior even at the small lattice sizes, as shown in Ref.~\cite{Bilgici:2009nm}.
In the case of $N_f=12$, we show the continuum extrapolation and scaling behavior for each step scaling in Fig.~\ref{fig:cont-lim}.
We found that the scaling behavior in the low energy region becomes worse.
The right figure in Fig.~\ref{fig:cont-lim} show the detailed behavior of the lowest energy, in which the input renormalized coupling is $u=2.36$.
We found that the data of $L/a=8$ is far from the data of the other lattice size and there is a large scaling violation in the step from $L/a=4$ to $sL/a=8$.

\begin{figure}[h]
\begin{center}
\rotatebox{0}{
   \includegraphics[width=6cm,clip]{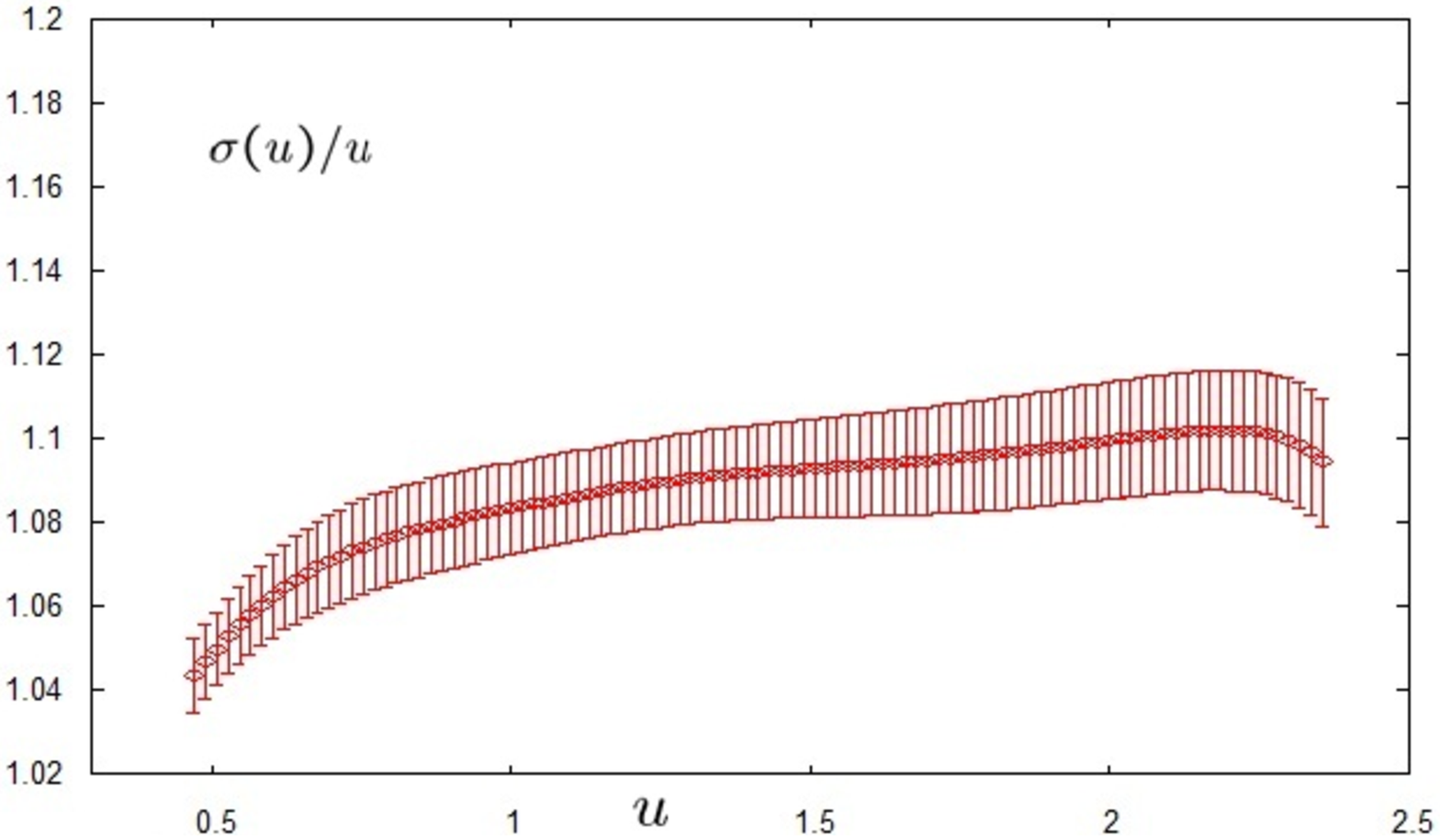}
\quad \quad
  \includegraphics[width=6cm,clip]{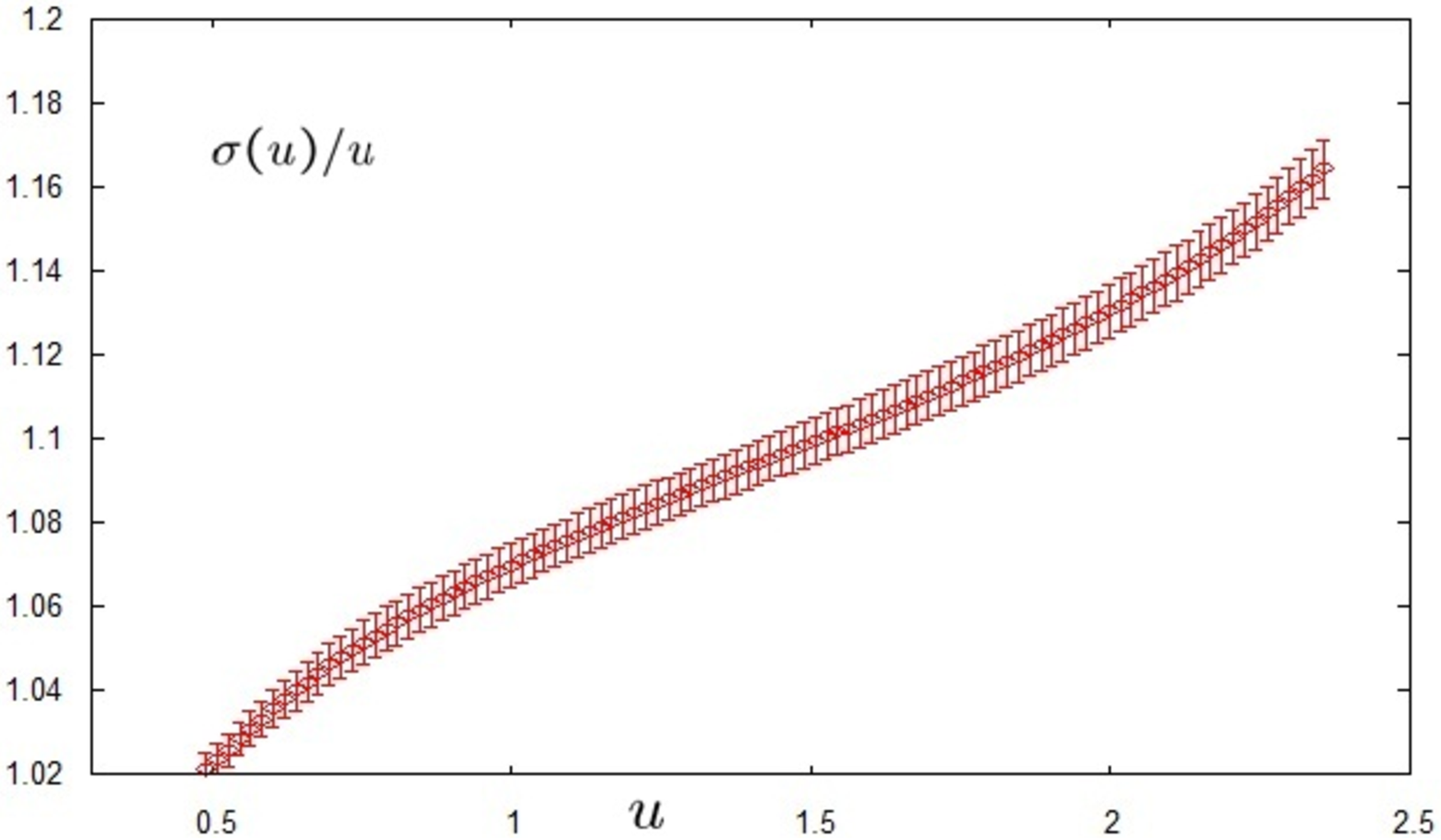}
}
  \caption{The growth rate of the scaling function using a linear extrapolation (the left figure) and a constant extrapolation (the right figure).}
  \label{fig:sigma-u}
\end{center}
\end{figure}
The growth rate of the scaling function ($\sigma(u)/u$) in the continuum limit is shown in Fig.~\ref{fig:sigma-u}, in which  we use a linear extrapolation of $(a/L)^2$ and a constant extrapolation.

The left figure in Fig.~\ref{fig:sigma-u} may show that the growth rate is decreasing for $u\gtrsim 2.2$.
However, the right one shows that there is no such signal.
Actually, the difference between these two extrapolations gives a systematic error of the renormalized coupling.
At the low energy region, there is more than $6\%$ systematic error.
The large systematic error comes from the large scaling violation of a small lattice size, as we show in Fig.~\ref{fig:cont-lim}.
To reduce the systematic errors and to conclude the existence of IR fixed point, we have to do some improvements or carry out the simulation for larger lattice size.

\section{Discussion and future direction}
In this paper, we investigate the growth rate of TPL coupling in low energy region $\beta \sim 5.0$.
The TPL coupling does not show the inversion of the order of lattice size even in the low $\beta$ region.
The lattice renormalized coupling has a discretization error and it depends on the renormalization scheme.
To remove the discretization effects and to estimate the systematic error, we took the continuum limit using two ways: a linear extrapolation of $(a/L)^2$ and a constant extrapolation.
The statistical error is less than $3\%$ even in a low energy region, but the systematic error is more than $6\%$.
We cannot give a conclusive statement for the existence of the fixed point due to the large systematic error.

To solve this difficulty and to obtain the conclusion of the fixed point search in $N_f=12$, there would be several directions to take.
One direction is to carry out the simulations on larger lattices.
In Ref.~\cite{SU2-ohki-paper} we carry out the step scaling using a set of large lattice size, $L/a=6,8,10,12$ and $sL/a=9,12,15,18$, for SU(2) eight flavor case.
In that study, the continuum extrapolation behaves nicely even in the low energy region.
Second direction may be given by some improvements of the action to reduce the discretized effect in the low energy.
Third one is to estimate the discretization error using the analytical calculation by lattice perturbation theory, and then to subtract this value of discretization error from the lattice raw data.
That may give a better scaling behavior.

Finally, we would like to show our future direction.
We have to measure the anomalous dimension of the field around the fixed point, if there is a fixed point in the infrared region.
The anomalous dimension of the operator on the fixed point is related with the conformal algebra of the conformal field theory, so that it should be scheme independent.
Comparision of its values which are measured in several schemes gives a conclusive statement of the existence of the fixed point.
Furthermore, from the phenomenological motivation we expect that the fermion composite operator has a large anomalous dimension and gives a origin of Higgs sector of standard model.
We will report a new method of measurement of the anomalous dimension in Ref.~\cite{SU3-final-paper}

\section*{Acknowledgements}
Numerical simulation was carried out on the vector supercomputer
NEC SX-8 at YITP, Kyoto University and at RCNP, Osaka University, and Hitachi SR11000 and IBM System Blue Gene Solution at KEK
under a support of its Large-Scale Simulation Program
(No. 09/10-22).

This work is supported in part of the Grant-in-Aid of the Ministry of Education (Nos. 
20105002,
20105005,
21105501,
21105508,
22740173,
and 21-897).
C.-J.D.~L. is supported by the National Science Council of Taiwan via grant 96-2122-M-009-020-MY3.

\end{document}